\documentclass[aps,reprint,twocolumn,notitlepage,numerical,superscriptaddress,nofootinbib]{revtex4-1}
\usepackage{amssymb, amsmath,framed,amsthm}

\makeatletter
\@ifpackageloaded{stix} 
{
}{                      
	\usepackage{mathrsfs}
	\usepackage{txfonts}              
}

\ifx\bm\undefined
  \newcommand{\bm}[1]{{\boldsymbol {\mathrm #1}}} 
\else
  \renewcommand{\bm}[1]{{\boldsymbol {\mathrm #1}}} 
\fi
\makeatother

\usepackage[unicode=true,pdfusetitle,bookmarks=true,bookmarksnumbered=false,bookmarksopen=false,breaklinks=false,pdfborder={0 0 0},backref=false,colorlinks=true]{hyperref}
\hypersetup{citecolor=blue,filecolor=blue,linkcolor=blue,urlcolor=blue}
\usepackage{color,soul}
\usepackage{graphicx}

\newcommand{\p}{\partial}
\newcommand{\f}[2]{\frac{#1}{#2}}
\newcommand{\mr}[1]{\mathrm{#1}}
\newcommand{\lf}{\left}
\newcommand{\ri}{\right}

\newcommand{\Div}{\nabla\cdot}
\newcommand{\Curl}{\nabla\times}
\newcommand{\nbl}{\nabla}
\newcommand{\para}{{||}}
\newcommand{\unit}[1]{\hat{\bm{#1}}}

\makeatletter
\newcommand{\vast}{\bBigg@{4}}
\newcommand{\Vast}{\bBigg@{5}}
\makeatother

\newcommand{\calM}{\mathcal{M}}

\DeclareMathAlphabet{\mathpzc}{OT1}{pzc}{m}{it}

\newcommand{\rmc}{\mathrm{c}}

\newcommand{\rme}{\mathrm{e}}

\newcommand{\rmi}{\mathrm{i}}

\newcommand{\rmA}{\mathrm{A}}

\newcommand{\rmF}{\mathrm{F}}

\newcommand{\rmS}{\mathrm{S}}

\begin{document}

\title{Modification of magnetohydrodynamic waves by the relativistic Hall effect}

\author{Yohei Kawazura}
\email{yohei.kawazura@physics.ox.ac.uk}
\affiliation{Rudolf Peierls Centre for Theoretical Physics, University of Oxford, Oxford OX1 3NP, UK}

\date{\today}
 
\begin{abstract}
  This study shows that a relativistic Hall effect significantly changes the properties of wave propagation by deriving a linear dispersion relation for relativistic Hall magnetohydrodynamics (HMHD). 
  Whereas, in non-relativistic HMHD, the phase and group velocities of fast magnetosonic wave become anisotropic with an increasing Hall effect, the relativistic Hall effect brings upper bounds to the anisotropies.
  The Alfv\'{e}n wave group velocity with strong Hall effect also becomes less anisotropic than non-relativistic case.
  Moreover, the group velocity surfaces of Alfv\'{e}n and fast waves coalesce into a single surface in the direction other than near perpendicular to the ambient magnetic field.
  It is also remarkable that a characteristic scale length of the relativistic HMHD depends on ion temperature, magnetic field strength, and density while the non-relativistic HMHD scale length, i.e., ion skin depth, depends only on density.
  The modified characteristic scale length increases as the ion temperature increases and decreases as the magnetic field strength increases.
\end{abstract}

\maketitle

\section{Introduction}
Plasmas are multiscale in nature, meaning that macroscopic dynamics are influenced by microscopic effects.
Although magnetohydrodynamics (MHD) is a simple and powerful model to capture ``overall'' dynamics in both space and laboratories~(see e.g., Refs.~\cite{Goedbloed2004,Goedbloed2010}), it fails to describe real phenomena when microscopic effects are not negligible.
To solve this issue, MHD has been extended by including various microscopic effects~\cite{Lust1959,Lighthill1960,Kimura2014,Lingam2015a}.
As one of the primary extensions, Hall magnetohydrodynamics (HMHD) was proposed \cite{Lighthill1960} and has been studied extensively in astrophysics, e.g., reconnection~\cite{Shay2001}, accretion disks~\cite{Wardle1999,Balbus2001,Lesur2014}, dynamo~\cite{Mininni2002,Lingam2015b}, non-linear Alfv\'{e}n wave~\cite{Mahajan2005}, and outflows~\cite{Lingam2016}, as well as in fusion~\cite{Mahajan1998,Hameiri2004}.

When studying astrophysical objects, it is also essential to consider relativistic effects.
Relativistic MHD~\cite{Lichnerowicz1967,Anile1990} has been a \textit{de facto} standard model for understanding large scale astrophysical phenomena. 
However, for the same reason as the non-relativistic MHD, the lack of microscopic effects may be a critical shortcoming.
Koide's extended version of the relativistic MHD (XMHD), in contrast, takes into account several microscopic effects originating from two-fluid nature~\cite{Koide2009,Koide2010}. 
This model has been highlighted in recent studies~\cite{Comisso2014,Asenjo2015,Asenjo2015a,Kawazura2017} 

To understand these various MHD models, it is crucial to consider the properties of linear wave propagation.
While the linear wave properties for non-relativistic ideal MHD have been widely known (see Refs.~\cite{Goedbloed2004,Goedbloed2010}), the detailed analysis of HMHD waves was not conducted until Hameiri et al. revealed that phase and group diagrams are deformed by the Hall effect \cite{Hameiri2005}.
In addition to the non-relativistic models, the dispersion relation for relativistic ideal MHD has been studied~\cite{Lichnerowicz1967,Anile1990,Kalra2000,Keppens2008}. 
Keppens and Meliani drew phase and group diagrams of the relativistic MHD showing that there is no qualitative difference between non-relativistic and relativistic diagrams in a fluid rest frame, except for the presence of the light limit \cite{Keppens2008}.
In addition to the relativistic MHD, there are studies on wave properties for relativistic electron-positron pair plasma~\cite{Soto2010, Asenjo2009,Munoz2014}.
For relativistic electron-ion plasma, whereas the dispersion relation for the relativistic XMHD was derived by Koide \cite{Koide2009} in specific wave vector configurations, a general dispersion relation in any wave vector direction has not been formulated.
Hence the phase and group diagrams for the relativistic XMHD are unknown.

One might assume that relativistic HMHD diagrams are similar to the non-relativistic HMHD diagrams, because the relativistic and the non-relativistic MHD diagrams are similar. 
In this paper, we show that it is not true; we formulate the linear dispersion relation of the relativistic HMHD in any wave vector direction and show that, depending on whether the Hall effect is relativistic or non-relativistic, there are differences in the way wave properties are changed.

\section{Derivation of relativistic HMHD dispersion relation}
Let us consider ion-electron plasma in Minkowski space-time with a metric $\mr{diag}(1,\, -1,\, -1,\, -1)$.
We begin with relativistic XMHD~\cite{Koide2009,Koide2010} which contains electron rest mass and thermal inertial effects, the thermal electromotive effect, and the Hall effect.
In this study, we focus on the Hall effect and ignore the other effects by assuming that the electron to ion mass ratio is zero and the electron temperature is at most on the order of the rest mass energy, i.e., $T_\rme/m_\rme c^2 \lesssim 1$ with electron's temperature $T_\rme$, rest mass $m_\rme$, and speed of light $c$.
From the latter condition, we may neglect all the terms including electron's thermal enthalpy and pressure in the momentum equation and the generalized Ohm's law~\cite{Note}. 
Thus we obtain the relativistic HMHD equations:
the continuity equation,
\begin{equation}
	\p_\nu\lf( nu^\nu \ri) = 0, 
	\label{e:HMHD continuity} 
\end{equation}
the momentum equation,
\begin{equation}
	\p_\nu\lf( nhu^\mu u^\nu \ri) = \p^\mu p + J_\nu F^{\mu\nu}, 
	\label{e:HMHD e.o.m.} 
\end{equation}
the generalized Ohm's law,
\begin{equation}
	enu_\nu F^{\mu\nu} - J_\nu F^{\mu\nu} = 0,
  \label{e:HMHD Ohm's law} 
\end{equation}
and the Maxwell's equations,
\begin{equation}
  \p_\mu F^{\mu\nu} = 4\pi J^\nu, \quad \p^\mu\lf( \epsilon_{\mu\nu\rho\sigma}{F}^{\rho\sigma} \ri) = 0.
	\label{e:Maxwell} 
\end{equation}
Multiplying $u_\mu$ by \eqref{e:HMHD e.o.m.}, the adiabatic equation is obtained,
\begin{equation}
	\p_\nu\lf(\sigma u^\nu \ri) = 0.
	\label{e:HMHD adiabaticity} 
\end{equation}
Here, $e$, $n$, $h$, $p$, and $\sigma$ are elementary charge, rest frame number density, ion thermal enthalpy, ion thermal pressure, and ion entropy density, respectively. 
We have also defined the four-velocity $u^\mu = (\gamma,\, \gamma\bm{v}/c)$, the Faraday tensor $F^{\mu\nu}$, the four-current $J^\nu = \p_\mu F^{\mu\nu} = (\rho_q,\, \bm{J}/c)$ where $\gamma = 1/\sqrt{1 - (|\bm{v}|/c)^2}$ is the Lorentz factor, $\epsilon_{\mu\nu\rho\sigma}$ is the four dimensional Levi-Civita symbol, and $\rho_q$ is the charge density. 
The only difference between the relativistic ideal MHD is the second term in \eqref{e:HMHD Ohm's law} which corresponds to a Hall term.
The 3+1 decompositions of (\ref{e:HMHD continuity})-(\ref{e:HMHD adiabaticity}) are written as
\begin{eqnarray*}
	\p_t\lf( n\gamma \ri) + \Div\lf( n\gamma\bm{v} \ri) = 0,
	\label{e:HMHD continuity 2} \\
	\p_t\lf( nh\gamma^2\bm{v} \ri) + \Div\lf( nh\gamma^2\bm{v}\bm{v} \ri) = -c^2\nbl p + c^2\rho_q\bm{E} + c\bm{J}\times\bm{B},
	\label{e:HMHD e.o.m. space component} \\
	c\lf( \gamma en - \rho_q \ri)\bm{E} + \lf( \gamma en\bm{v} - \bm{J} \ri)\times\bm{B} = 0,
	\label{e:HMHD Ohm's law space component} \\
	\p_t\lf( \sigma\gamma \ri) + \Div\lf( \sigma\gamma\bm{v} \ri) = 0, \\
	\label{e:HMHD adiabaticity 2} 
  \Div\bm{E} = 4\pi\rho_q,
	\label{e:Maxwell Div E} \\
  -\p_t\bm{E} + c\Curl\bm{B} = 4\pi\bm{J},
	\label{e:Maxwell dt E} \\
	\Div\bm{B} = 0,
	\label{e:Maxwell Div B} \\
  \p_t\bm{B} + c\Curl\bm{E} = 0,
	\label{e:Maxwell dt B} 
\end{eqnarray*}
with the electric field $\bm{E}$ and the magnetic field $\bm{B}$.
Here, the time components of \eqref{e:HMHD e.o.m.} and \eqref{e:HMHD Ohm's law} are not shown since the former is dependent to the adiabatic equation \eqref{e:HMHD adiabaticity}, and the latter is dependent to the spatial component of the Ohm's law itself.

Next, we linearize these equations by separating the variables into homogeneous background fields denoted by subscripts 0 and small amplitude perturbations denoted by tilde symbols that are proportional to $\exp(i\bm{k}\cdot\bm{x} - \omega t)$ with the wave vector $\bm{k}$ and the frequency $\omega$.
We set the frame as the fluid rest frame by assuming $\bm{v}_0 = 0$ (the frame may be Lorentz transformed to the lab frame in the same manner as \cite{Keppens2008}). 
The background part of Maxwell's equations lead $\rho_{q0} = 0, \; \bm{E}_0 = 0$, and $\bm{J}_0 = 0$.
In the following, we assume the ideal gas equation of state $h = mc^2 + [\Gamma/(\Gamma - 1)]T$ where $T$ is the ion temperature, and $\Gamma = 4/3$ is a specific heat ratio in ultra-relativistic case~\cite{Taub1948}.
Then we get a set of equations that the perturbations satisfy,
\begin{eqnarray}
  \label{e:RHMHD continuity Fourier}
  -i\omega \tilde{n} + i n_0 \bm{k}\cdot\tilde{\bm{v}} = 0, \\
  \label{e:RHMHD e.o.m. Fourier}
  -i\omega n_0h_0\tilde{\bm{v}} = -i\bm{k}c^2\tilde{p} + c\tilde{\bm{J}}\times\bm{B}_0, \\
  \label{e:RHMHD Ohm's law Fourier}
  \tilde{\bm{E}} + \f{1}{c}\lf( \tilde{\bm{v}} - \f{\tilde{\bm{J}}}{en_0} \ri)\times\bm{B}_0 = 0, \\
  \label{e:RHMHD adiabaticity Fourier}
  -i\omega\tilde{p} = -\f{\Gamma p_0}{n_0}i\omega\tilde{n}, \\
  \label{e:RHMHD Ampere Fourier}
  \f{4\pi}{c}\tilde{\bm{J}} = i\bm{k}\times\tilde{\bm{B}} + \f{1}{c}i\omega\tilde{\bm{E}}, \\
  \label{e:RHMHD divB Fourier}
  i\bm{k}\cdot\tilde{\bm{B}} = 0, \\
  \label{e:RHMHD Faraday's law Fourier}
  i\bm{k}\times\tilde{\bm{E}} = \f{1}{c}i\omega\tilde{\bm{B}},
\end{eqnarray}
where $h_0$ is the backgrout part of $h$, and $p_0 = n_0 T_0$ is the equilibrium ion pressure.
We note that the previous study on the relativistic XMHD wave~\cite{Koide2009} assumed $\tilde{\rho}_q = 0$, which results in $\bm{k}\cdot\tilde{\bm{E}} = \bm{k}\cdot\tilde{\bm{J}} = 0$.
Although this assumption makes the analysis simple (see Appendix \ref{s:Koide's dispersion}), it is not generally true. 
In the present study, we do not assume this condition.
Below, we omit the tilde symbols.

Without loss of generality, one may assume $\bm{k} = (k_\perp,\, 0,\, k_\para)$ and $\bm{B}_0 = (0,\, 0,\, B_0)$.
We ignore the entropy mode, i.e., $\omega \ne 0$.
Manipulating \eqref{e:RHMHD continuity Fourier}--\eqref{e:RHMHD Faraday's law Fourier}, we obtain the dispersion relation (see Appendix \ref{s:derivation} for the detailed derivation).
\begin{eqnarray}
  &&(\omega^2 - k_\para^2 V_\rmA^2)\times\nonumber\\
  &&\Bigg\{ \omega^4 - \lf[ k^2\lf(V_\rmA^2 + \f{n_0 h_0}{\calM}C_\rmS^2\ri) + c^{-2}C_\rmS^2V_\rmA^2k_\para^2 \ri]\omega^2 + k^2k_\para^2 V_\rmA^2 C_\rmS^2 \Bigg\} \nonumber \\
  && = \delta_\rmi^2 V_\rmA^2\omega^2 c^{-4}\lf( \omega^2 - k_\para^2c^2 \ri)\lf( \omega^2 - k^2c^2 \ri)\lf( \omega^2 - k^2 C_\rmS^2 \ri)
  \label{e: RHMHD dispersion relation}
\end{eqnarray}
where
\begin{eqnarray}
  \calM = n_0 h_0 + \f{B_0^2}{4\pi}, \; \f{C_\rmS}{c} = \sqrt{\f{\Gamma p_0}{n_0 h_0}}, \; \f{V_\rmA}{c} = \f{B_0}{\sqrt{4\pi \calM}}, \; \delta_\rmi^2 = \f{h_0^2}{4\pi\calM e^2} \nonumber\\
  \label{e:params}
\end{eqnarray}
are total (fluid and magnetic) enthalpy~\cite{Keppens2008}, sound speed, Alfv\'{e}n speed, and \textit{modified} ion skin depth, respectively.
We also have an identity $1 - V_\rmA^2/c^2 = n_0h_0/\calM$.
Taking the $\delta_\rmi \to 0$ limit, \eqref{e: RHMHD dispersion relation} becomes the relativistic ideal MHD dispersion relation~\cite{Keppens2008}.  
In the non-relativistic limit, $\delta_\rmi$ becomes the familiar ion skin depth $d_\rmi = \sqrt{mc^2/4\pi n_0 e^2}$.
One finds that $\delta_\rmi$ depends on both $B_0$ and $T_0$ as well as $n_0$.
This is remarkably different from the non-relativistic ion skin depth $d_\rmi$ which depends only on $n_0$.
$\delta_\rmi$ monotonically increases with increasing $T_0$ and monotonically decreases with increasing $B_0$.   
The $T_0$ dependence is straightforward, i.e., a high temperature induces large effective mass resulting in a long inertial length.
This dependence has been pointed out in past studies~\cite{Comisso2014, Kawazura2017, Mahajan2016}.
However, the shrink of the inertial length by large $B_0$ is not trivial.
Let us use $\beta = 8\pi n_0 T_0/B_0^2$ and $\hat{T} = T_0/mc^2$ as parameters instead of $T_0$ and $B_0$. 
One may rewrite the modified ion skin depth as
\begin{equation}
  \delta_\rmi^2 = d_\rmi^2\f{\hat{h}^2}{\hat{h} + 2\hat{T}/\beta},
  \label{e:delta_i}
\end{equation}
where $\hat{h} = h_0/mc^2$. 
One finds that $\delta_\rmi$ vanishes for small beta plasma.
Therefore if the magnetic field is very strong, the plasma behaves like ideal MHD.
One may interpret this as follows.
The Alfv\'{e}n speed is written as $V_\rmA = \Omega_\rmc \delta_\rmi$ where $\Omega_\rmc = ceB_0/h_0$ is the relativistic cyclotron frequency. 
When one takes the non-relativistic limit (shown below), this becomes the familiar expression  $V_\rmA = \Omega_\rmc d_\rmi$ with the non-relativistic cyclotron frequency $\Omega_\rmc = eB_0/mc^2$.
Whereas the cyclotron frequency is proportional to the magnetic field strength, the Alfv\'{e}n speed may not exceed the speed of light (see \eqref{e:params}); hence the modified skin depth is required to shrinks as the magnetic field strength increases. 

We note that the dispersion relation \eqref{e: RHMHD dispersion relation} and the modified inertial length $\delta_\rmi$ are valid as long as the relativistic two-fluid equations is correct for ion-electron plasma since (\ref{e:HMHD continuity})-(\ref{e:HMHD adiabaticity}) are rigorously derived by the relativistic two-fluid equations~\cite{Koide2009,Koide2010}.
However, it is proven that there is limitations for non-relativistic HMHD dispersion relation when it is derived from a kinetic theory~\cite{Ito2004,Hirose2004}.
It is a open question whether the dispersion relation \eqref{e: RHMHD dispersion relation} is derived from the relativistic kinetic theory.

Next let us consider the non-relativistic limit,
\begin{eqnarray}
  h_0 \to mc^2,\;\; \calM \to n_0mc^2, \;\; \f{\omega^2}{c^2k_\para^2} \to 0, \;\; \f{\omega^2}{c^2k^2} \to 0.
\end{eqnarray}
We get $\delta_\rmi \to d_\rmi$, $(V_\rmA/c)^2 \to 2\hat{T}/\beta$, and $(C_\rmS/c)^2 \to \Gamma\hat{T}$.
Then \eqref{e: RHMHD dispersion relation} yields non-relativistic HMHD dispersion relation~\cite{Hameiri2005},
\begin{eqnarray}
  (\omega^2 - k_\para^2 V_\rmA^2)\lf\{ \omega^4 - k^2\lf( V_\rmA^2 + C_\rmS^2 \ri)\omega^2 + k^2k_\para^2 V_\rmA^2 C_\rmS^2 \ri\} \nonumber\\
  = d_\rmi^2 V_\rmA^2\omega^2 k_\para^2 k^2\lf( \omega^2 - k^2 C_\rmS^2 \ri)
  \label{e: HMHD dispersion relation}
\end{eqnarray}
Since in the non-relativistic case, $C_\rmS/V_\rmA$ does not depend on $\hat{T}$, the shape of the phase and group diagrams are independent of $\hat{T}$~\cite{Goedbloed2004}. 

\section{Analysis of relativistic HMHD dispersion relation}
In the beginning, we show two apparent differences between the relativistic dispersion relation~\eqref{e: RHMHD dispersion relation} and the non-relativistic one~\eqref{e: HMHD dispersion relation}.
First, for exactly perpendicular propagation, viz., $k_\para = 0$, the right hand side of~\eqref{e: RHMHD dispersion relation} is finite whereas the right hand side of~\eqref{e: HMHD dispersion relation} vanishes.
Therefore, for this direction, the non-relativistic Hall effect does not change the wave properties~\cite{Hameiri2005}, but the relativistic Hall effect does.
Second, the right hand side of \eqref{e: RHMHD dispersion relation} is quartic with respect to $\omega^2$ while the left hand is cubic. 
This means that the relativistic HMHD has one additional wave solution that neither appears in relativistic ideal MHD nor non-relativistic HMHD (the right hand side of~\eqref{e: HMHD dispersion relation} is quadratic).
As we show below, this extra wave is super luminous which becomes light wave at infinitely large $k\delta_i$.
The other three waves are shear Alfv\'{e}n wave, and slow and fast magnetosonic waves.
Below, for notational brevity, superscripts A, F, and S denote the Alfv\'{e}n, fast, and slow waves, respectively.

Let us start by the analysis in $k_\para = 0$ direction.
Again, the non-relativistic dispersion relation~\eqref{e: HMHD dispersion relation} becomes ideal MHD in this direction.
The relativistic dispersion relation~\eqref{e: RHMHD dispersion relation} becomes the quadratic equation.
The solutions are analytically obtained as
\begin{widetext}
\begin{equation}
  \lf( v_\mr{ph}^\pm \ri)^2 = \f{1}{2}\lf\{  1 + \hat{C}_\mr{S}^2 + \f{1}{(k_\perp\delta_\rmi)^2 \hat{V}_\rmA^2} \pm \sqrt{\lf[ 1 + \hat{C}_\mr{S}^2 + \f{1}{(k_\perp\delta_\rmi)^2 \hat{V}_\rmA^2} \ri]^2 - 4\lf[ \hat{C}_\rmS^2 + \f{1}{(k_\perp\delta_\rmi)^2 \hat{V}_\rmA^2}\lf( \hat{V}_\rmA^2 + \f{n_0 h_0}{\calM}C_\rmS^2 \ri) \ri] } \ri\},
  \label{e:perpendicular solution}
\end{equation}
\end{widetext}
where $\bm{v}_\mr{ph} = (v_{\mr{ph}\perp},\, 0,\, v_{\mr{ph}||}) = (\omega/ck)\,\bm{n}$ is the normalized phase velocity with $\bm{n} = \bm{k}/k$, $\hat{V}_\rmA = V_\rmA/c$ and $\hat{C}_\rmS = C_\rmS/c$ which are normalized Alfv\'{e}n and sound speed.
One may show that $v_\mr{ph}^+$ ($v_\mr{ph}^-$) is always larger (smaller) than unity.
Thus $v_\mr{ph}^+$ is super luminous wave.
These two solutions become $(v_\mr{ph}^+)^2 \to 1$ and $(v_\mr{ph}^-)^2 \to \hat{C}_\rmS^2$ for $k_\perp\delta_\rmi \to \infty$ limit, and $(v_\mr{ph}^+)^2 \to \infty$ and $(v_\mr{ph}^-)^2 \to \hat{V}_\rmA^2 + \hat{C}_\rmS^2(n_0 h_0/\calM) = \hat{C}_\rmS^2 + (1 - \hat{C}_\rmS^2)\hat{V}_\rmA^2$ (which is the fast wave phase speed for the ideal MHD) for $k\delta_\rmi \to 0$ limit.
The behavior of the super luminous solution is the same as the ordinary wave in the strongly magnetised relativistic electron-positron plasma~\cite{Barnard1986,Lyutikov1998}.
The disappeared two solutions become the shear Alfv\'{e}n and the slow waves in $k_\para \ne 0$ direction.

Next we consider rough dependence of phase speed on $k\delta_\rmi$ when the magnitude of $k\delta_\rmi$ is large. 
As shown in Ref.~\cite{Hameiri2005} for the non-relativistic case, the phase speed of the three HMHD waves are $v_\mr{ph}^\rmF \sim O((kd_\rmi)^2)$, $v_\mr{ph}^\rmS \sim O(1/(kd_\rmi)^2)$, and $v_\mr{ph}^\rmA \sim O(1/(kd_\rmi)^2)$, respectively; 
since the left hand side in~\eqref{e: HMHD dispersion relation} is $\sim v_\mr{ph}^6$ and the right hand side is $\sim (kd_\rmi)^2v_\mr{ph}^4$, the solution with $O((kd_\rmi)^2)$ exists.
Thus, $v_\mr{ph}^\rmF$ increases as $kd_\rmi$ increases.
On the other hand for the relativistic case~\eqref{e: RHMHD dispersion relation}, the left hand side is $\sim v_\mr{ph}^6$ and the right hand side is $\sim (k\delta_\rmi)^2v_\mr{ph}^8$. 
Therefore, a solution with $O((k\delta_\rmi)^2)$ does not exist, and the phase speed of the all wave may not increase as $k\delta_\rmi$ increases. 
This fact is reasonable for the fast, slow, and Alfv\'{e}n waves because their phase speed may not exceed $c$.
In $k_\para = 0$ direction, this is exactly confirmed by \eqref{e:perpendicular solution} which is principally $O(1/(k_\perp\delta_\rmi)^2)$ and decreases monotonically.

As shown in Ref.~\cite{Hameiri2005}, in non-relativistic case, $v_\mr{ph}^\rmF$ increases except in $k_\para = 0$ direction as $kd_\rmi$ increases.
Since the diagram for the ideal MHD fast wave is a circular shape, it becomes a dumbbell shape stretched in the parallel direction at large $kd_\rmi$.
On the other hand, in the relativistic case, $v_\mr{ph}^\rmF$ may not increase;  
especially in $k_\para = 0$ direction, $(v_\mr{ph}^\rmF)^2$ decreases from $\hat{C}_\rmS^2 + (1 - \hat{C}_\rmS^2)\hat{V}_\rmA^2$ to $\hat{C}_\rmS^2$ as $k\delta_\rmi$ increases.
Therefore the resulting phase diagram at large $k\delta_\rmi$ is less anisotropic than the non-relativistic phase diagram.

Next we consider the aforementioned non-relativistic limit more carefully.
There are two relativistic effects included, the ion thermal inertia effect and the displacement current.
The former effect is eliminated by assuming $\hat{T} \ll 1$, which results in $h_0 \to mc^2$.
The situation is divided in two cases depending on the value of $\hat{T}/\beta$ since $\calM \to n_0 mc^2 (1 + 2\hat{T}/\beta)$.
For a moderately or weakly magnetised case, we get $\calM \to n_0mc^2$ which results in $\delta_\rmi \to d_\rmi$.
Therefore the change of the inertial length does not happen.
For a strongly magnetised case, we get $\delta_\rmi^2 \to d_\rmi^2/(1 + 2\hat{T}/\beta)$; thus the inertial length may change, and the change is due to the displacement current.
In both cases, the structure of the dispersion relation does not change from \eqref{e: RHMHD dispersion relation}.
Therefore, the super luminous solution still exists, and the phase speed for all waves is $O(1/(k\delta_\rmi)^2)$.

The other relativistic effect, displacement current, is eliminated by taking $\calM \to n_0mc^2, \; \omega^2/c^2k_\para^2 \to 0$, and $\omega^2/c^2k^2 \to 0$.
Accordingly we get $\delta_\rmi \to d_\rmi$, $(V_\rmA/c)^2 \to 2\hat{T}/\beta$, and $(C_\rmS/c)^2 \to \Gamma\hat{T}$.
Obviously, this limit prohibits the super luminous solution; in fact, the right hand side of \eqref{e: HMHD dispersion relation} is $\omega^4$.
Since the order of the right hand side has been lowered, phase speed with $O((k\delta_\rmi)^2)$ is allowed. This solution is a non-relativistic fast wave.

In summary, among the two relativistic effects, the ion thermal inertia only contributes to the change of the inertial length, and the displacement current contributes to the emergence of the super luminous solution and the wave dependence on $k\delta_\rmi$.
As we show in the next section, this relativistic wave dependence on $k\delta_\rmi$ caused by the displacement current changes the phase and group diagrams dramatically.

Next we graphically show the phase diagram for the specific plasma parameter.
For a given $\beta$, $\hat{T}$, and $kd_\rmi$, the phase velocity $\bm{v}_\mr{ph}$ is determined as a function of $\theta = \cos^{-1}(k_\para/k)$ by solving \eqref{e: RHMHD dispersion relation}. 
Both $C_\rmS/c$ and $V_\rmA/c$ are monotonically increasing functions of $\hat{T}$ with upper bounds of $\sqrt{\Gamma - 1}$ and $\sqrt{2(\Gamma-1)/[\beta\Gamma + 2(\Gamma - 1)]}$, respectively. 
Since $C_\rmS$ and $V_\rmA$ almost become the upper bound values for $\hat{T} \gtrsim 1$, we consider only a $\hat{T} = 1$ case.
Since the relativistic Hall effect disappears in very low beta plasmas as mentioned above, we consider $\beta = 0.1$ case. 
Such plasma parameters are relevant to Poynting flux dominated gamma ray bursts (see for example~\cite{Zhang2010}).
These settings yield $\delta_\rmi/d_\rmi = 1$.

\begin{figure*}
  \centering
  \includegraphics*[width=1.0\textwidth]{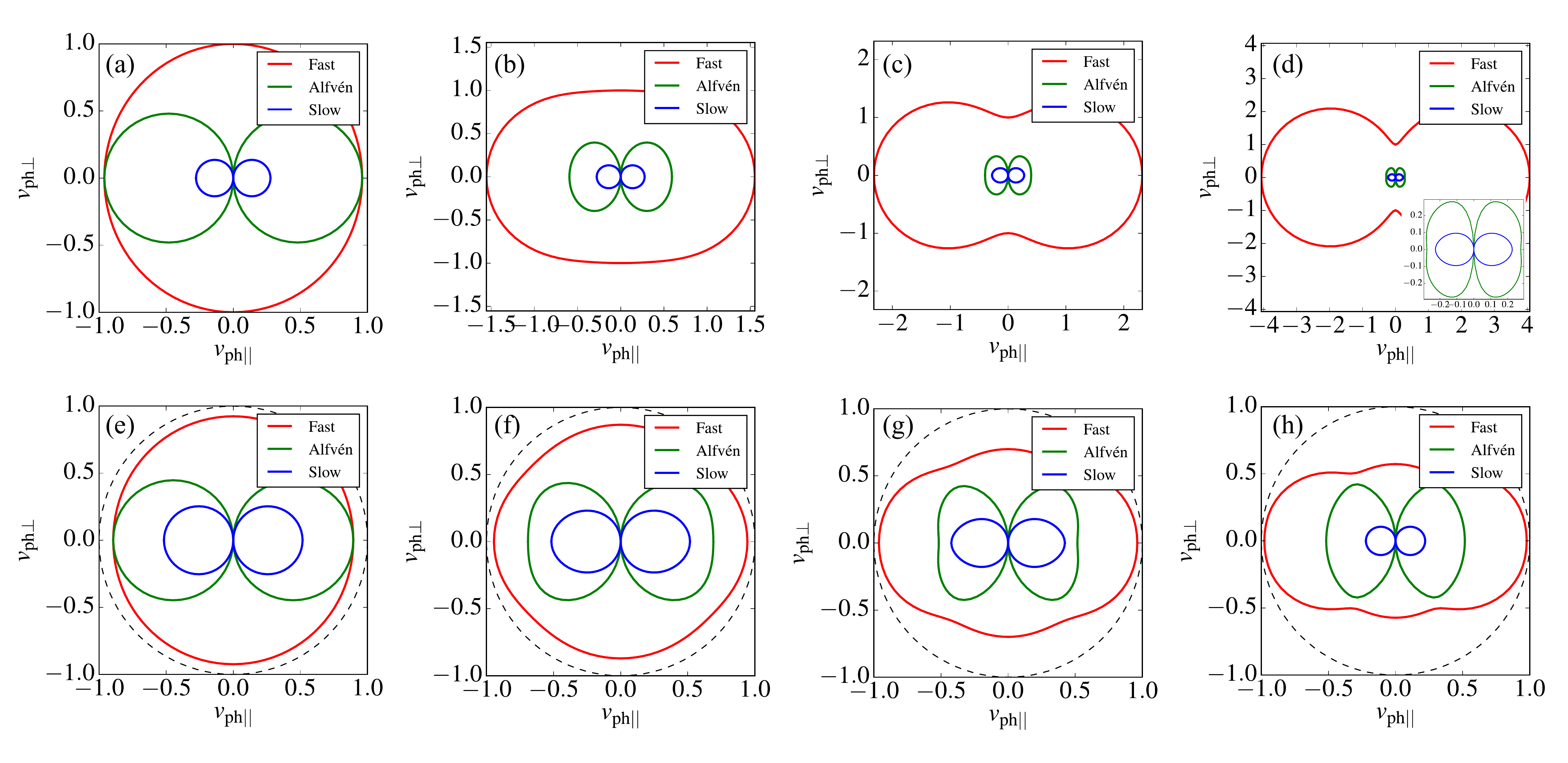}
  \caption{The phase diagram for (a) non-relativistic ideal MHD, (b)-(d) non-relativistic HMHD, (e) relativistic ideal MHD, and (f)-(h) relativistic HMHD. $\beta$ and $\hat{T}$ are fixed at 0.1 and 1.0, and $kd_\rmi$ varies as $(0.0, 1.0, 2.0, 4.0)$ from left to right panels. The vertical and horizontal axes in (a)--(d) are normalized by $v_\mr{ph}^\rmF$ with $kd_\rmi = 0$ and $\theta = 0$. The inset figure in (d) is an enlargement near the Alfv\'{e}n wave. The red, green, and blue curves indicate the fast, Alfv\'{e}n, and the slow modes, respectively. The broken circles indicate the light speed.}
  \label{f:phase_diagram}
\end{figure*}
Figure~\ref{f:phase_diagram} shows the phase diagrams with various $kd_\rmi = (0,\, 0.5,\, 1.0,\, 4.0)$ for (a)-(d) non-relativistic and (e)-(h) relativistic cases.
For the non-relativistic cases, the shape of the diagram is independent from $\hat{T}$, hence $\bm{v}_\mr{ph}$ is normalized by $v^\rmF_{\mr{ph}}$ at $\theta = \pi/2$, which is the same for any $d_\rmi$.
The non-relativistic diagrams Figs.~\ref{f:phase_diagram} (a)-(d) are the same as those in Ref.~\cite{Hameiri2005}.
We find the phase speed of the fast wave ($v^\rmF_\mr{ph}$) increases with increasing $kd_\rmi$ except in the $\theta = \pi/2$ direction, 
hence the circular shaped phase velocity surface for $kd_\rmi = 0$ (Fig.~\ref{f:phase_diagram} (a)) becomes elliptic (Fig.~\ref{f:phase_diagram} (c)) and finally dumbbell-like in shape (Fig.~\ref{f:phase_diagram} (d)).
This means that the fast wave becomes anisotropic in small scale.
Another observation is that the phase speeds of the Alfv\'{e}n wave ($v^\rmA_\mr{ph}$) and slow wave ($v^\rmS_\mr{ph}$) decrease with increasing $kd_\rmi$.
Thus, the fast wave gets separated from the other two waves. 
This separation is appreciable especially in the parallel direction.

Let us compare these results with the relativistic HMHD diagrams (Figs.~\ref{f:phase_diagram} (e)-(h)).
Whereas the relativistic ideal MHD diagram (Fig.~\ref{f:phase_diagram} (e)) is qualitatively the same as the non-relativistic ideal MHD one (Fig.~\ref{f:phase_diagram} (a)), the relativistic HMHD diagrams (Figs.~\ref{f:phase_diagram} (f)-(h)) are significantly different from those of the non-relativistic HMHD (Figs.~\ref{f:phase_diagram} (b)-(d)).
The anisotropy of the fast wave is moderated so that the shape of the phase velocity surface does not become a dumbbell shape.
This isotropization is explained as follows.
In non-relativistic HMHD, the anisotropy is created by the selective increase in $v^\rmF_{\mr{ph}\para}$ as $kd_\rmi$ increases.
In relativistic HMHD, on the other hand, $v^\rmF_{\mr{ph}}$ may not exceed the light limit, hence there is no room for the significant increase in $v^\rmF_{\mr{ph}\para}$.
Meanwhile, Figs.~\ref{f:phase_diagram} (e)-(h) show that $v^\rmF_{\mr{ph}\perp}$ decreases as $kd_\rmi$ increases (recall that $v^\rmF_{\mr{ph}\perp}$ at $\theta = \pi/2$ is changeable for increasing $kd_\rmi$ unlike non-relativistic HMHD). 
However, $v^\rmF_{\mr{ph}\perp}$ will eventually reach $v^\rmA_{\mr{ph}\perp}$ because $v^\rmA_{\mr{ph}\perp}$ does not decrease. 
Since $v^\rmF_{\mr{ph}\perp}$ may not be smaller than $v^\rmA_{\mr{ph}\perp}$, the decrease in $v^\rmF_{\mr{ph}\perp}$ is saturated at some value of $kd_\rmi$.
In short, $v^\rmF_{\mr{ph}}$ is bounded from above by the light limit and from below by $v^\rmA_{\mr{ph}}$. 
Thus the anisotropy will stop increasing at some $kd_\rmi$.

\begin{figure}
  \centering
  \includegraphics*[width=0.5\textwidth]{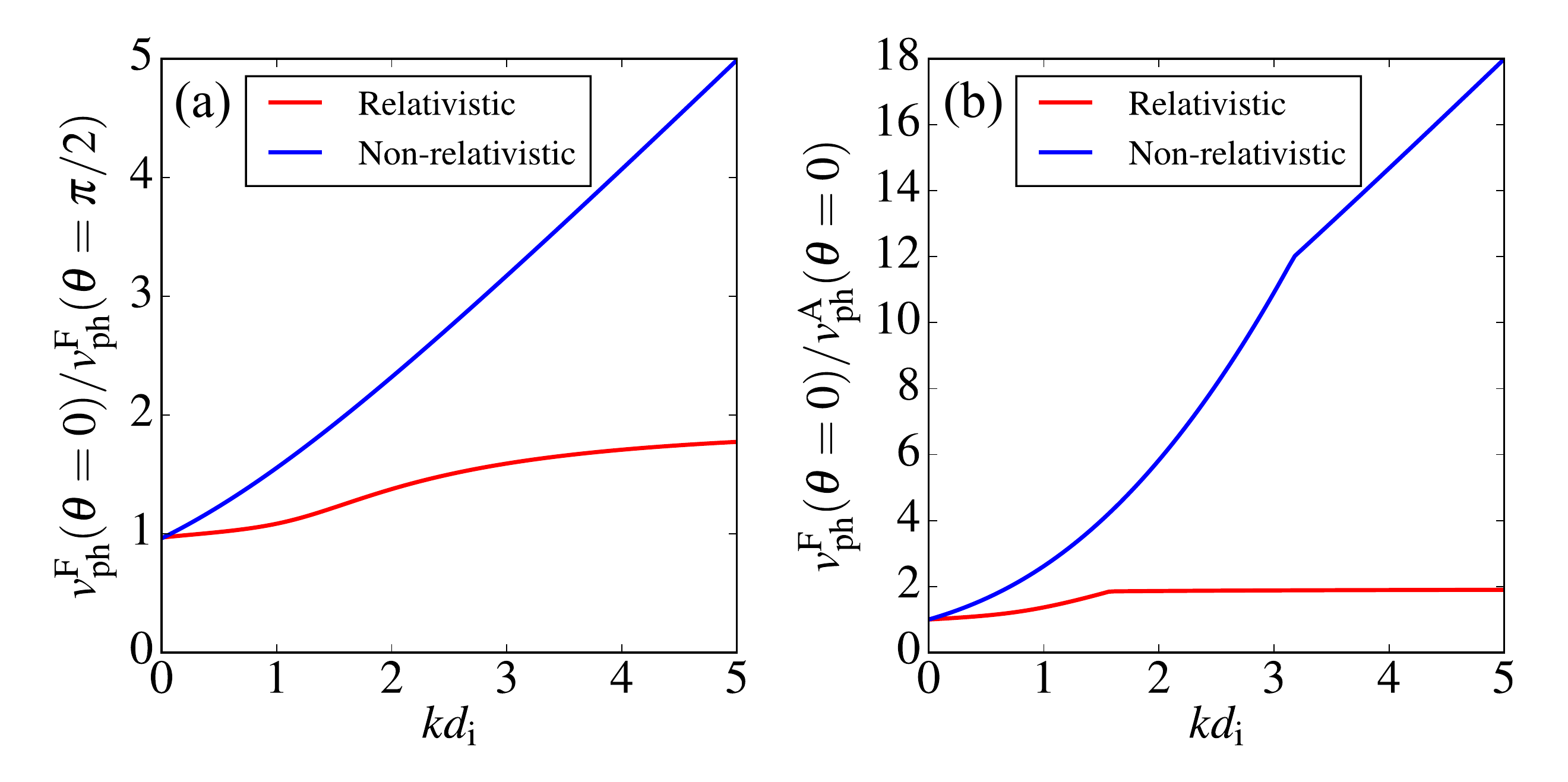}
  \caption{(a) The anisotropy of the fast wave phase velocity and (b) the ratio of the fast phase speed to the Alfv\'{e}n phase speed at $\theta = 0$. The red and blue curves indicate the relativistic and the non-relativistic cases, respectively. $\beta$ and $\hat{T}$ are fixed at 0.1 and 1.0.}
  \label{f:phase_di_dependence}
\end{figure}
Figure~\ref{f:phase_di_dependence} explicitly illustrates this scenario. Figure~\ref{f:phase_di_dependence} (a) shows a measure of the fast wave anisotropy defined by $v^\rmF_\mr{ph}(\theta=0)/v^\rmF_\mr{ph}(\theta=\pi/2)$ as a function of $kd_\rmi$ for the non-relativistic and relativistic cases.
One finds that the anisotropy increases almost linearly in the non-relativistic case whereas it is bounded by $\sim 2$ in the relativistic case. 
Figure~\ref{f:phase_di_dependence} (b) shows the ratio of the fast phase speed to the Alfv\'{e}n phase speed at $\theta = 0$.
While, in the non-relativistic case, the difference between the two speeds increases endlessly, the fast wave phase speed becomes at most twice as fast as the Alfv\'{e}n phase speed for $kd_\rmi \gtrsim 1.5$ in the relativistic case.
For the relativistic case, since the maximum of $v^\rmA_\mr{ph\perp}$ is almost the same as $v^\rmA_\mr{ph}(\theta = 0)$ for large $kd_\rmi$ (see Fig.~\ref{f:phase_diagram}(h)), the lower bound of $v^\rmF_\mr{ph\perp}$ is almost the same as $v^\rmA_\mr{ph}(\theta = 0)$.
Therefore $v^\rmF_\mr{ph}(\theta = 0)/v^\rmA_\mr{ph}(\theta = 0) \sim 2$ corresponds to the fast wave anisotropy of $\sim 2$.

Next we consider a normalized group velocity $\bm{v}_\mr{gr} = (v_{\mr{gr}\perp},\, 0,\, v_{\mr{gr}||}) = c^{-1}\p\omega/\p\bm{k}$, see Appendix \ref{s:group velocity} for full expression.
\begin{figure*}
  \centering
  \includegraphics*[width=1.0\textwidth]{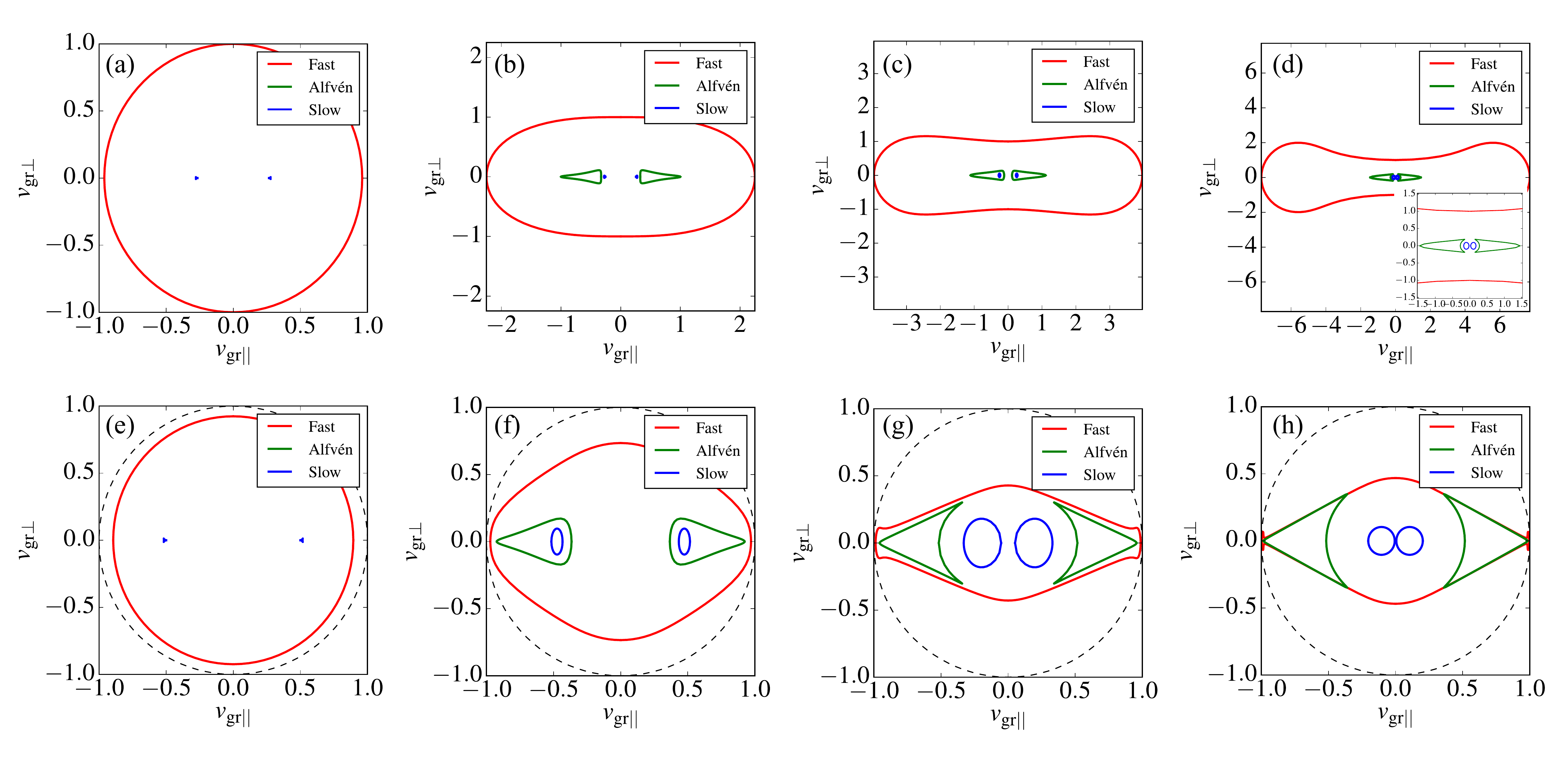}
  \caption{The group diagram for (a) non-relativistic ideal MHD, (b)-(d) non-relativistic HMHD, (e) relativistic ideal MHD, and (f)-(h) relativistic HMHD. $\beta$ and $\hat{T}$ are fixed at 0.1 and 1.0, and $kd_\rmi$ varies as $(0.0, 1.0, 2.0, 4.0)$ from left to right panels. The vertical and horizontal axes in (a)--(d) are normalized by $v_\mr{gr}^\rmF$ with $kd_\rmi = 0$ and $\theta = 0$. The inset figure in (d) is an enlargement near the Alfv\'{e}n wave. The broken circles indicate the light speed.}
  \label{f:group_diagram}
\end{figure*}
Figure~\ref{f:group_diagram} shows the group diagrams for (a)-(d) non-relativistic and (e)-(h) relativistic cases with the same parameters as~Fig.~\ref{f:phase_diagram}.
The non-relativistic diagrams Figs.~\ref{f:group_diagram} (a)-(d) are the same as those in Ref.~\cite{Hameiri2005}.
Here, we again find that there is no qualitative difference between relativistic ideal MHD (Fig.~\ref{f:group_diagram} (e)) and non-relativistic ideal MHD (Fig.~\ref{f:group_diagram} (a)).
On the other hand, the relativistic HMHD diagrams (Figs.~\ref{f:group_diagram} (f)-(h)) are significantly different from those of the non-relativistic HMHD (Figs.~\ref{f:group_diagram} (b)-(d)).
In the non-relativistic case, the behavior of the fast wave group velocity ($\bm{v}^\rmF_\mr{gr}$) is similar to that of phase velocity; in the beginning, the group velocity surface is circular, then it becomes elliptic and successively dumbbell-like as $kd_\rmi$ increases. 
Since Alfv\'{e}n wave becomes dispersive when the Hall effect is present, its group velocity becomes a triangle.
We find that the Alfv\'{e}n group velocity surface becomes acute-angled triangle at large $kd_\rmi$ (see the inset of Fig.~\ref{f:group_diagram} (d)).

Let us compare these behaviors with the relativistic HMHD (Figs.~\ref{f:group_diagram} (f)-(h)).
We find that the fast and Alfv\'{e}n group velocity surfaces coalesce into a single surface at large $kd_\rmi$.
In the relativistic case, $v^\rmF_\mr{gr\para}$ and $v^\rmA_\mr{gr\para}$ is not allowed to increase as $kd_\rmi$ increases since they are almost at the light limit for $kd_\rmi = 0$ (Fig.~\ref{f:group_diagram} (e)).
On the other hand, $v^\rmF_\mr{gr\perp}$ decreases as $kd_\rmi$ increases.
Thus, the fast wave group velocity surface tends to be elliptic (Fig.~\ref{f:group_diagram} (e) $\to$ (g)).
Meanwhile, $v^\rmA_\mr{gr\perp}$ increases as $kd_\rmi$ increases (Fig.~\ref{f:group_diagram} (e) $\to$ (g)).
At some point, the increasing $v^\rmA_\mr{gr\perp}$ becomes the same value as the decreasing $v^\rmF_\mr{gr\perp}$.
Thus the coalesce of the fast and Alfv\'{e}n waves is realized.
Since the fast wave speed may not be smaller than the Alfv\'{e}n wave speed, the fast wave diagram no longer becomes dumbbell-like in shape.
The Alfv\'{e}n wave diagram also becomes less anisotropic since only $v^\rmA_\mr{gr\perp}$ increases as $kd_\rmi$ increases.

\begin{figure}
  \centering
  \includegraphics*[width=0.5\textwidth]{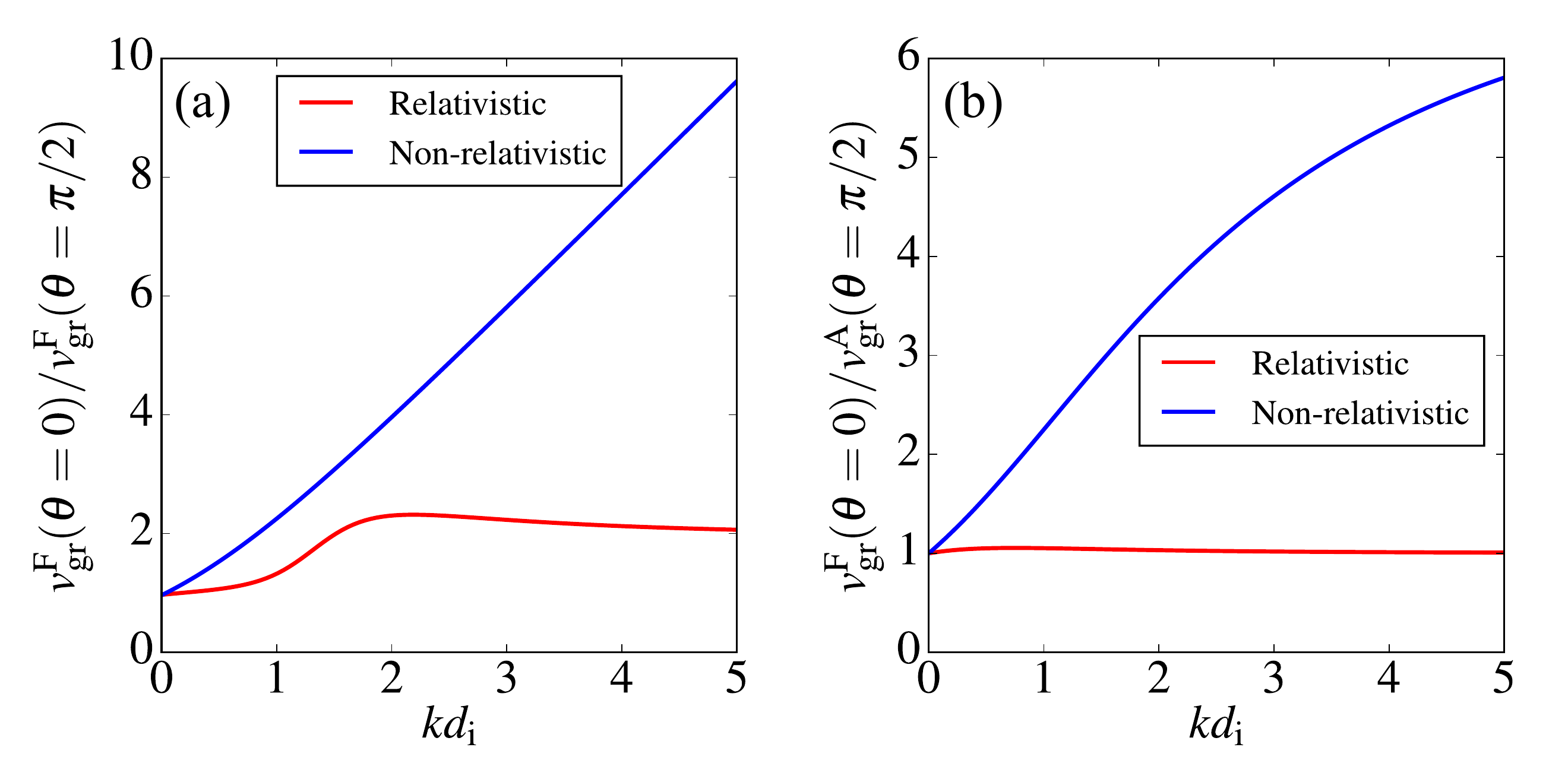}
  \caption{(a) The anisotropy of the fast wave group velocity and (b) the ratio of the fast group speed at $\theta = 0$ to the Alfv\'{e}n group speed at $\theta = \pi/2$. The red and blue curves indicate the relativistic and the non-relativistic cases, respectively. $\beta$ and $\hat{T}$ are at fixed at 0.1 and 1.0.}
  \label{f:group_di_dependence}
\end{figure}
Figure~\ref{f:group_di_dependence} (a) shows a measure of the fast wave anisotropy defined by $v^\rmF_\mr{gr}(\theta=0)/v^\rmF_\mr{gr}(\theta=\pi/2)$ as a function of $kd_\rmi$ for the non-relativistic and relativistic cases.
One finds that the anisotropy linearly increases in the non-relativistic case while it saturates at $\sim 2$ in the relativistic case.
Figure.~\ref{f:group_di_dependence} (b) shows the ratio of the maximum of $v^\rmF_\mr{gr}$ to the maximum of $v^\rmA_\mr{gr}$, i.e.,  $v^\rmF_\mr{gr}(\theta=0)/v^\rmA_\mr{gr}(\theta=\pi/2)$.
Here, $v^\rmA_\mr{gr}(\theta=\pi/2)$ corresponds to the vertex of the Alfv\'{e}n surface on the $v_{\mr{gr}\perp}$ axis.
Whereas the fast and Alfv\'{e}n waves separate in the non-relativistic case, such a separation is prohibited in the relativistic case.

\section{Conclusions}
We have shown that the relativistic Hall effect changes the MHD wave properties in a different way from the non-relativistic Hall effect; namely, \textit{isotropization} and \textit{coalescence} of the fast and shear Alfv\'{e}n waves.
It is also remarkable that the characteristic scale length $\delta_\rmi$ depends on ion temperature, magnetic field strength as well as density.
This is different from the non-relativistic ion skin depth $d_\rmi$ which depends only on density.
The modified ion inertial length increases as the ion temperature increases whereas it decreases as the magnetic field strength increases.
Therefore the Hall effect disappears, and plasma behaves like ideal MHD in very strong magnetic field.

\section*{Acknowledgements} 
This work was supported by STFC grant ST/N000919/1 and JSPS KAKENHI Grant 26800279.

\appendix

\section{Derivation of \eqref{e: RHMHD dispersion relation}} \label{s:derivation}
In this section, we explicitly derive the relativistic HMHD dispersion relation \eqref{e: RHMHD dispersion relation}.
From \eqref{e:RHMHD continuity Fourier} and \eqref{e:RHMHD adiabaticity Fourier}, we get
\begin{eqnarray*}
  p = \f{n_0 h_0}{\omega}\f{C_\rmS^2}{c^2}(\bm{k}\cdot\bm{v}).
\end{eqnarray*}
Substituting this into \eqref{e:RHMHD e.o.m. Fourier}, we get
\begin{equation}
  -\omega n_0 h_0 \bm{v} = -\f{n_0 h_0}{\omega}C_\rmS^2(\bm{k}\cdot\bm{v})\bm{k} + \f{c}{i}\bm{J}\times\bm{B}_0.
  \label{e:RHMHD e.o.m.2 Fourier}
\end{equation}
$\bm{k}\times\eqref{e:RHMHD Ohm's law Fourier}$ yields 
\begin{equation}
  \omega\bm{B} + k_\para B_0\bm{v} - (\bm{k}\cdot\bm{v})\bm{B}_0 - \f{1}{en_0}\lf[ k_\para B_0\bm{J} - (\bm{k}\cdot\bm{J})\bm{B}_0 \ri] = 0.
  \label{e:RHMHD induction Fourier}
\end{equation}
$\bm{k}\cdot\eqref{e:RHMHD Ampere Fourier}$ yields 
\begin{eqnarray}
  \bm{k}\cdot\bm{J} = \f{B_0}{4\pi c}i \omega\lf[ -(\bm{k}\times\bm{v})\cdot\unit{z} + \f{i}{4\pi e n_0 c}\lf( \omega^2 - c^2 k^2 \ri)B_z \ri].
  \label{e:RHMHD k . J}
\end{eqnarray}
$\bm{k}\cdot\eqref{e:RHMHD e.o.m.2 Fourier}$ is manipulated using $\eqref{e:RHMHD Ampere Fourier}$ and \eqref{e:RHMHD Faraday's law Fourier} as
\begin{equation}
  \bm{k}\cdot\bm{v} = -\f{B_0 \omega\lf(\omega^2 - c^2k^2\ri)B_z}{4\pi n_0 h_0\lf(\omega^2 - C_\rmS^2 k^2\ri)}
  \label{e:RHMHD k . v}
\end{equation}
Eliminating $v_z$ from the $z$ components of \eqref{e:RHMHD e.o.m.2 Fourier} and \eqref{e:RHMHD induction Fourier} and using \eqref{e:RHMHD k . v}, we get
\begin{equation}
  \f{\omega}{k_\para B_0}B_z = -\f{\omega^2 - C_\rmS^2 k_\para^2}{\omega k_\para}\f{B_0\lf(\omega^2 - c^2 k^2\ri)B_z}{4\pi n_0 h_0\lf(\omega^2 - C_\rmS^2 k^2\ri)} + \f{1}{en_0}\lf( J_z - \f{\bm{k}\cdot\bm{J}}{k_\para} \ri).
  \label{e:RHMHD k . J and J_z}
\end{equation}
Substituting \eqref{e:RHMHD k . J} into $(\bm{k}\times\eqref{e:RHMHD e.o.m.2 Fourier})\cdot\unit{z}$, we get
\begin{equation}
  (\bm{k}\times\bm{v})\cdot\unit{z} = i \f{cB_0}{\omega\calM}\lf[ k_\para J_z + \f{B_0\omega}{(4\pi c)^2en_0}\lf(\omega^2 - c^2k^2\ri)B_z \ri].
  \label{e:RHMHD k x v . z}
\end{equation}
Then we back-substitute this into \eqref{e:RHMHD k . J} to get
\begin{equation}
  \bm{k}\cdot\bm{J} = \f{V_\rmA^2}{c^2}k_\para J_z + \lf( \f{V_\rmA^2}{c^2} - 1 \ri)\f{\omega\lf(\omega^2 - c^2k^2\ri)B_0 B_z}{(4\pi c)^2 en_0}.
  \label{e:RHMHD k . J2}
\end{equation}
Next we substitute \eqref{e:RHMHD k x v . z} and \eqref{e:RHMHD k . J2} into $(\bm{k}\times\eqref{e:RHMHD induction Fourier})\cdot\unit{z}$ to get
\begin{eqnarray}
  &&\lf[ \f{4\pi V_\rmA^2}{c}\omega k_\para^2 - 4\pi c\omega k^2 - \f{cB_0^2}{\omega\calM}k_\para^2\lf(\omega^2 - c^2k^2\ri) \ri]J_z = \nonumber \\
  &&\lf[ -\lf( \f{V_\rmA^2}{c^2} - 1 \ri)\f{\omega^2 k_\para \lf(\omega^2 - c^2k^2\ri)B_0}{4\pi cen_0} + \f{B_0^3}{(4\pi)^2 cen_0\calM}k_\para\lf(\omega^2 - c^2k^2\ri)^2 \ri.\nonumber\\ 
  && \lf. - \f{k_\para B_0}{4\pi n_0 ce}\lf(\omega^2 - c^2k^2\ri)^2 \ri]B_z.
  \label{e:RHMHD Jz1}
\end{eqnarray}
Substituting \eqref{e:RHMHD k . J2} into \eqref{e:RHMHD k . J and J_z}, we get
\begin{eqnarray}
  &&-\f{1}{en_0}\lf( \f{V_\rmA^2}{c^2} - 1 \ri)J_z = \lf[ \f{\omega}{k_\para B_0} + \f{B_0}{4\pi n_0 h_0}\f{\lf(\omega^2 - C_\rmS^2 k_\para^2\ri)\lf(\omega^2 - c^2 k^2\ri)}{\omega k_\para\lf(\omega^2 - C_\rmS^2 k^2\ri)} \ri.\nonumber\\  
  &&\lf.+ \lf( \f{V_\rmA^2}{c^2} - 1 \ri)\f{\omega\lf(\omega^2 - c^2k^2\ri)B_0}{(4\pi c en_0)^2k_\para} \ri]B_z.
  \label{e:RHMHD Jz2}
\end{eqnarray}
Finally, we obtain the dispersion relation \eqref{e: RHMHD dispersion relation} by eliminating $J_z$ from \eqref{e:RHMHD Jz1} and \eqref{e:RHMHD Jz2}.

\section{Reduction to the dispersion relation in Ref.~\cite{Koide2009}} \label{s:Koide's dispersion}
In Ref.~\cite{Koide2009} the dispersion relation for XMHD is derived in the specific condition.
The wave vector and the perturbed velocity are set in the configuration $\bm{k} = k_x \unit{x}$ and $\bm{v} = v_x\unit{x}$.
Furthermore, rather restrictive condition $\bm{k}\cdot\bm{E} = \bm{k}\cdot\bm{J} = 0$, which is led by the assumption $\rho_q = 0$, is imposed.
Under these conditions, we may simplify \eqref{e:RHMHD e.o.m.2 Fourier} and \eqref{e:RHMHD induction Fourier} as
\begin{equation}
  -\omega n_0 h_0 v_x = -\f{n_0 h_0}{\omega}C_\rmS^2k_x^2 v_x + \f{c}{i}J_y B_0
\end{equation}
and
\begin{equation}
  \omega B_z - k_x v_x B_0 = 0.
\end{equation}
Combining with \eqref{e:RHMHD Ampere Fourier} and \eqref{e:RHMHD Faraday's law Fourier}, we obtain the very simplified dispersion relation,
\begin{equation}
  \omega^2 = \f{4\pi n_0 h_0 C_\rmS^2 + c^2 B_0^2}{4\pi n_0 h_0 + B_0^2}k_\perp^2.  
\end{equation}
This is the dispersion relation for the fast wave in Ref.~\cite{Koide2009}.

\section{Group velocity} \label{s:group velocity}
In a similar manner to Ref.~\cite{Hameiri2005}, straightforward algebraic manipulation of \eqref{e: RHMHD dispersion relation} yields 
\begin{equation}
  \bm{v}_\mr{gr} = \f{\bm{K} + (k\delta_\rmi)^2\bm{L}}{M + (k\delta_\rmi)^2N}
  \label{e: HMHD group velocity}
\end{equation}
with
\begin{widetext}
\begin{eqnarray}
  \bm{K} &=& \lf\{ \lf( \hat{V}_\rmA^2 + \f{n_0 h_0}{\calM}\hat{C}_\rmS^2 \ri)v_\mr{ph}^4 - \lf[ \hat{V}_\rmA^2 + \lf( 1 + \f{n_0 h_0}{\calM} \ri)\hat{C}_\rmS^2 \ri]\hat{V}_\rmA^2\lf( \f{k_\para}{k} \ri)^2v_\mr{ph}^2 + \hat{C}_\rmS^2\hat{V}_\rmA^4\lf( \f{k_\para}{k} \ri)^4 \ri\}\bm{n} \nonumber\\
  & & + \lf\{ \lf( 1 + \hat{C}_\rmS^2 \ri)\hat{V}_\rmA^2\lf( \f{k_\para}{k} \ri)v_\mr{ph}^4 - \lf[ 2\lf( \f{k_\para}{k} \ri)^2\hat{V}_\rmA^2 \hat{C}_\rmS^2 + \lf[ \hat{V}_\rmA^2 + \lf( 1 + \f{n_0 h_0}{\calM} \ri)\hat{C}_\rmS^2 \ri] \ri]\hat{V}_\rmA^2\lf( \f{k_\para}{k} \ri)v_\mr{ph}^2 + 2\hat{C}_\rmS^2\hat{V}_\rmA^4\lf( \f{k_\para}{k} \ri)^3 \ri\}\bm{b} \nonumber\\
  \nonumber\\
  \bm{L} &=& \hat{V}_\rmA^2\lf\{ -\lf( 1 + \hat{C}_\rmS^2 \ri)v_\mr{ph}^4 + \lf[ \lf( 1 + \hat{C}_\rmS^2 \ri)\lf( \f{k_\para}{k} \ri)^2 + 2\hat{C}_\rmS^2 \ri]v_\mr{ph}^2 - 2\hat{C}_\rmS^2\lf( \f{k_\para}{k} \ri)^2 \ri\}v_\mr{ph}^2\bm{n} + \hat{V}_\rmA^2\lf\{ -\lf( \f{k_\para}{k} \ri)v_\mr{ph}^4 + \lf( 1 + \hat{C}_\rmS^2 \ri)\lf( \f{k_\para}{k} \ri)v_\mr{ph}^2 - \hat{C}_\rmS^2\lf( \f{k_\para}{k} \ri) \ri\}v_\mr{ph}^2\bm{b} \nonumber\\
  \nonumber\\
  M &=& 3v_\mr{ph}^5 - 2\lf[ \lf( 1 + \hat{C}_\rmS^2 \ri)\lf( \f{k_\para}{k} \ri)^2\hat{V}_\rmA^2 + \lf( \hat{V}_\rmA^2 + \f{n_0 h_0}{\calM}\hat{C}_\rmS^2 \ri) \ri]v_\mr{ph}^3 + \lf\{ \lf( \f{k_\para}{k} \ri)^2\hat{V}_\rmA^2\hat{C}_\rmS^2 + \lf[ \hat{V}_\rmA^2 + \lf( 1 + \f{n_0 h_0}{\calM} \ri)\hat{C}_\rmS^2 \ri] \ri\}\hat{V}_\rmA^2\lf( \f{k_\para}{k} \ri)^2v_\mr{ph} \nonumber\\
  \nonumber\\
  N &=& \hat{V}_\rmA^2\lf\{ 4v_\mr{ph}^6 - 3\lf[ 1 + \hat{C}_\rmS^2 + \lf( \f{k_\para}{k} \ri)^2 \ri]v_\mr{ph}^4 + 2\lf[ \lf( 1 + \hat{C}_\rmS^2 \ri)\lf( \f{k_\para}{k} \ri)^2 + \hat{C}_\rmS^2 \ri]v_\mr{ph}^2 - \hat{C}_\rmS^2\lf( \f{k_\para}{k} \ri)^2 \ri\}v_\mr{ph},
\end{eqnarray}
\end{widetext}
with $\bm{b} = \bm{B}_0/B_0$.

\bibliographystyle{apsrev4-1}
\bibliography{references}

\end{document}